\documentclass{PoS}

\title{Life and Death of Very Massive Stars}

\ShortTitle{Life and Death of the Most Massive Stars}

\author{\speaker{Norhasliza Yusof}$^a$, {Hasan Abu Kassim}$^{a}$, Raphael Hirschi$^{bc}$, Paul. A. Crowther$^{d}$, Olivier Schnurr$^{de}$, Richard J. Parker$^{df}$ and Simon P. Goodwin$^{d}$\\
\llap{$^a$}Department of Physics, University of Malaya, 50603 Kuala Lumpur, Malaysia\\
\llap{$^b$}Astrophysics Group, pdfAM, University of Keele, Lennard-Jones Lab, Keele, ST5 5BG, UK\\
\llap{$^c$}Institute for the Physics and Mathematics of the Universe, University of Tokyo, 5-1-5 Kashiwanoha, Kashiwa, 277-8583, Japan\\
\llap{$^d$}Department of Physics and Astronomy, University of Sheffield, Sheffield S3 7RH, UK \\
\llap{$^e$}Astrophysikalisches Institut of Postdam, An der Sternwarte 16, D-14482 Postdam, Germany \\
\llap{$^f$}Institute for Astronomy, ETH Zurich, Wolfgang-Pauli-Strasse 27, 8093 Zurich, Switzerland\\

E-mail: \email{norhasliza@siswa.um.edu.my}}

\abstract{We recently determined the mass of the most massive star known to the date, R136a1 with a mass at birth 320 times the mass of our sun, as well as the mass of several other stars that are more massive than 150 M$_\odot$. Such massive stars ($\sim$150-300 M$_\odot$) may end their life as pair-instability supernovae (PISN) if they retain enough mass until they die. We have calculated a grid of stellar evolution models in order to investigate the impact of mass loss and rotation on the evolution and fate of these very massive stars. As mass loss is very strong at solar metallicity, our models predict that
most of the very massive stars will die as type Ic SNe. Only slowly and non-rotating stars
at metallicities below that of the LMC might retain enough mass to produce a
PISN. This would mean that the first stellar generations might have produced
PISN although their chemical signature is not observed in extremely metal poor
stars in the halo of our galaxy.}

\FullConference{11th Symposium on Nuclei in the Cosmos, NIC XI\\
        July 19-23, 2010\\
        Heidelberg, Germany}

\begin{document}

\section{Introduction}

The fate of a star is determined by its mass, composition, rotation rate and binarity, the mass being the most important parameter. Massive stars with an initial mass between
8 and 140 M$_\odot$ will evolve through all burning stages and form an iron core. 
The fate of the star follows this sequence as a function of increasing initial mass: SNII--SNIb--SNIc--BH without SN. 
For initial masses between 140 and 260 M$_\odot$, the stars are thought to die as
pair-instability supernovae if they do not lose much mass during their evolution, which is expected for low-metallicity non-rotating stars \cite{Heger02,Langer07}. 
PISNe are very bright and produce large quantities of iron \cite{Heger02}. Even if they are rare, they contribute significantly to the chemical evolution of galaxies. Their existence is debated because their chemical signature is not observed in extremely metal poor stars \cite{Umeda}. The absence of this chemical signature could be due to the fact that stars more massive than 140 M$_\odot$ do not form. This argument was supported by local observations. Indeed, the initial mass function (IMF) of the most massive young cluster in the Galaxy measured using HST seems to indicate a lack of stars with mass more massive than 150 M$_\odot$ \cite{Figer05}. 
However, observations of 2007bi \cite{GalYam} indicate that it is consistent with PISN theoretical models. 

In this work, we present evidence supporting the existence of stars more massive than 150 M$_\odot$. 
We also investigate the effect of
rotation, mass loss and metallicity on
the evolution of very massive stars. Main sequence evolutionary models are compared with the observations and we find an excellent agreement for the NGC 3603 and R136 systems \cite{pac10}. Finally, we discuss the fate of these very massive stars. 




\section{Stellar Evolution Models}
A grid of stellar evolution model for 85, 120, 150, 200, 300 and
500 M$_\odot$ stars has been calculated until at least the end of
the main sequence using the Geneva stellar evolution code
\cite{Eggenberger} both with and without rotation. These models include the physics of rotation and mass loss,
which are both very important in the evolution of very massive stars.
The models include all the effects of the rotation: centrifugal
support, mass-loss enhancement and especially mixing in radiative
zones \cite{Eggenberger}. In these models, we compute the main
sequence evolution using theoretical mass loss
rates for O type star \cite{Vink01}, which match
observations very well \cite{Mokiem07}. 
We consider that the star enters the Wolf-Rayet phase when the surface hydrogen
content $X_H < 30\%$ if T$_{\rm eff}$ $\geq$  10,000 K, during which an
empirical mass-loss calibration is followed \cite{Nugis00}. We
choose the ratio of the initial velocity to critical rotation of
$v_{init}/v_{crit}$= 0.4 for the rotating  models, which corresponds
to surface equatorial velocities around 350 km$^{-1}$ for 85
M$_\odot$ model and 450 km$^{-1}$ for the 500 M$_\odot$ model.

\begin{figure*}[h!]
\centering
\includegraphics[width=1.0\textwidth]{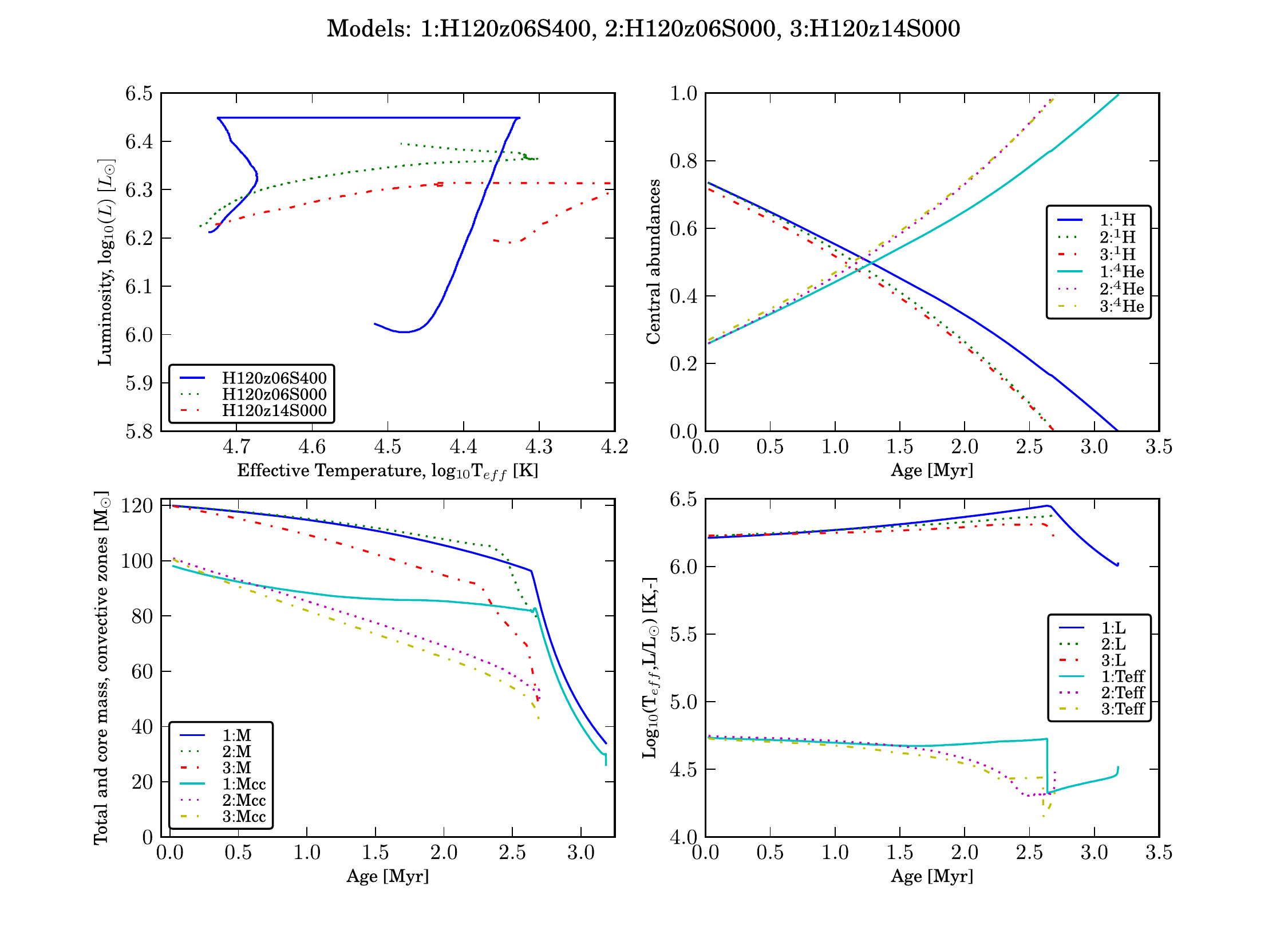}
\caption{Comparison between main sequence evolutionary predictions for both rotating (Z = 0.006: H120z06S400) 
and non-rotating (Z = 0.006:H120z06S000, Z = 0.014: H120z14S000) 120 M$_\odot$ models. 
For model H120z06S400, the horizontal lines in the top left-hand panel correspond to the transition to the WR phase, phase during which the photosphere is in the wind rather than at the surface of the star.}
\label{figure1}
\end{figure*}

The evolution tracks of 120 M$_\odot$ models are presented in Fig. 1. In
the figure we can see that the effect of rotation is significant.
Additional mixing above the H-burning convective core causes helium to be mixed up
to the surface. Thus the opacity in the outer layers decreases and the surface of rotating models remains hotter than that of  non-rotating
models. During the main sequence, the effective temperature of the rotating 
models remains hot, around
45 000-55 000 K whereas for the non-rotating models the effective
temperature decreases steadily to 20 000-25 000 K. Hence, rapidly rotating
stars evolve directly to the classically W-R phase while the slow
rotator stars are expected to become $\eta$ Carinae-like blue
luminous variable stars (see Meynet and Meader 2005 \cite{Meynet05}). In Fig. 1, we compare solar metallicity with LMC
metallicity models and we can see that lower metallicity models reach higher
luminosities. This is due to the lower mass loss.

\begin{figure*}[h]
\begin{center}
\includegraphics[width=.6\textwidth,angle=270,clip]{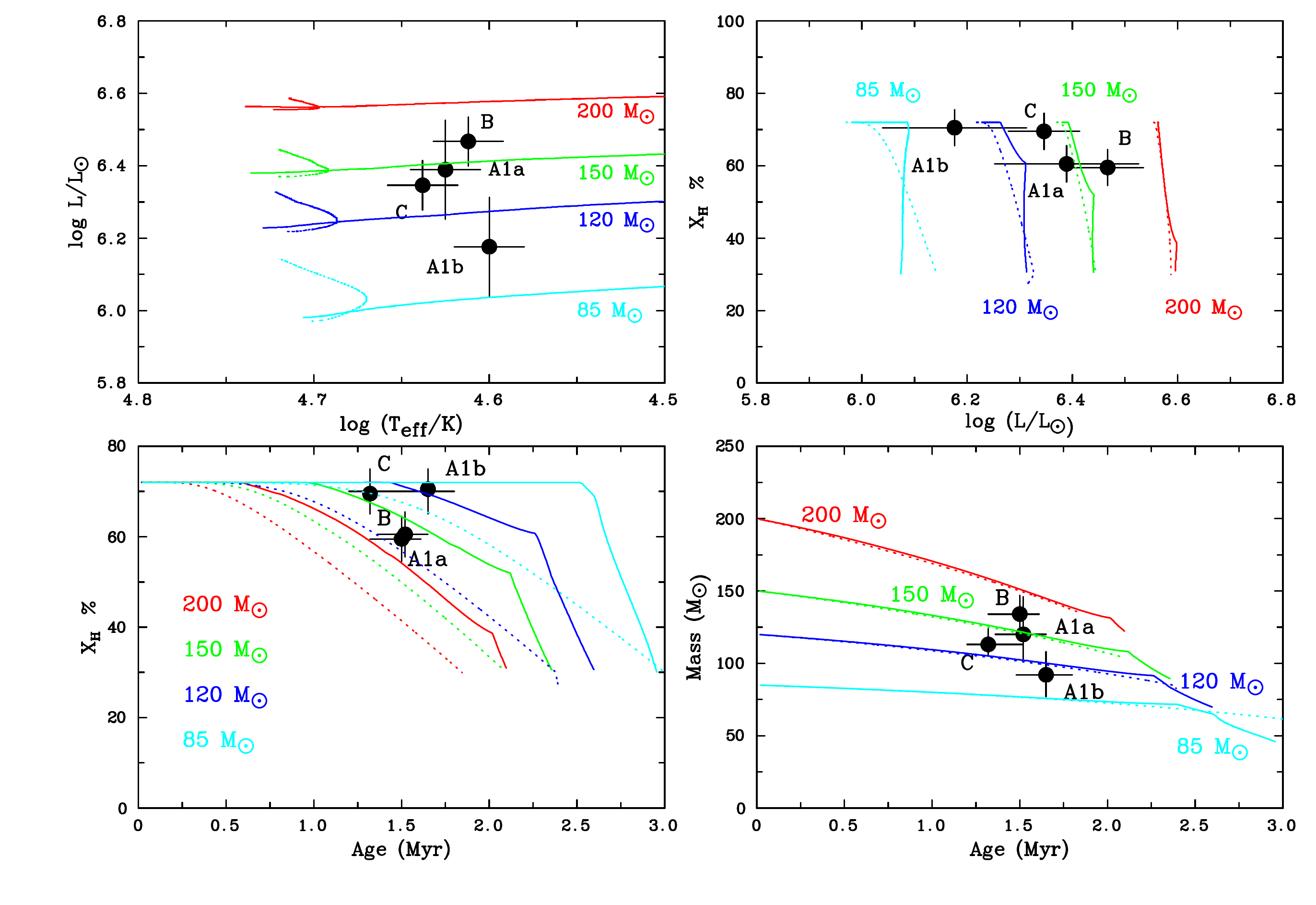}

\caption{Comparison between Solar metallicity (Z = 1.4\%) models
calculated for the main-sequence evolution of 85 - 200 M$_{\odot}$ stars
(initially rotating at $V_{\rm init}/v_{\rm crit}$ =0.4 [dotted] and 0
[solid]), and the
physical properties  derived from  spectroscopic analysis of NGC 3603 WN6h
stars. }
\label{ngc3603_evol}
\end{center}
\end{figure*}

\begin{figure*}[h]
\begin{center}
\includegraphics[width=0.6\textwidth,angle=270,clip]{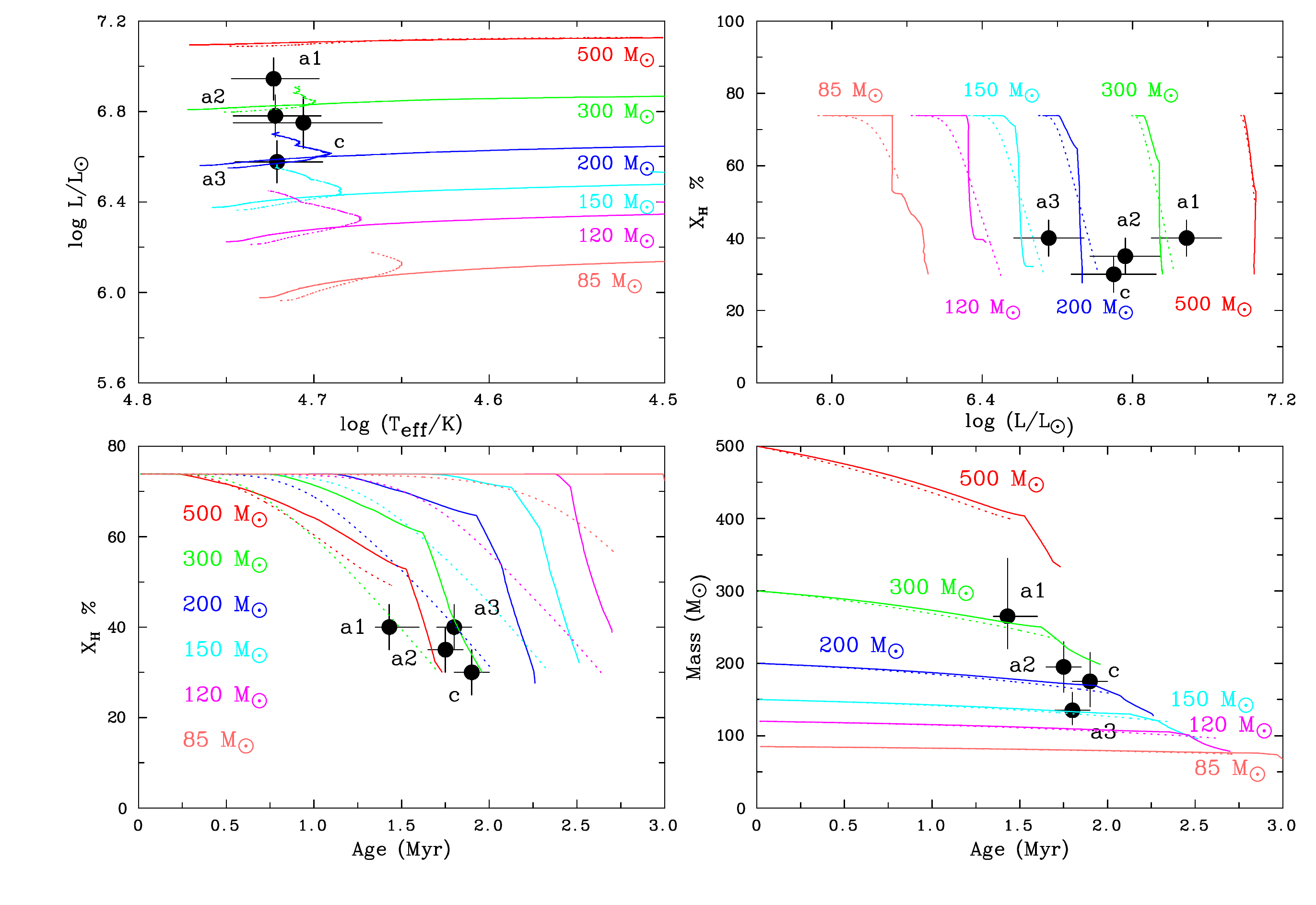}
\caption{Comparison between LMC-metallicity models calculated for
the main-sequence evolution of 85 - 500 M$_{\odot}$ stars,
initially rotating at $v_{\rm init}/v_{\rm crit}$ = 0.4 (dotted)
or 0 (solid) and the physical properties derived from our
spectroscopic analysis. } \label{r136_fig1to4}
\end{center}
\end{figure*}

\section{Mass Determination of the Most Massive Stars}
The results from the evolutionary models were compared with spectroscopic analysis of two young star clusters,
i.e. NGC 3603 and R136. Details of the spectroscopic analysis of the star clusters can be found in Crowther et al. \cite{pac10}. In Fig. 2, we compare the evolutionary models of solar metallicity with the
 observational properties of NGC 3603 WN stars while in Fig. 3, the evolutionary models of LMC metallicity
 are compared to the properties of R136 stars.

For NGC 3603, we find a good agreement with dynamical masses for A1a and
A1b for initially non-rotating models at ages of ~1.5 $\pm$ 0.1  Myr. 
This provides strong support for our mass determination method (for more details see  Crowther et al. 2010 \cite{pac10}). For R136, we obtain good agreement between the R136 stars and initially rapidly rotating, 165 -- 320 M$_{\odot}$ models at 
ages of $\sim$1.7 $\pm$ 0.2 Myr. In particular, we determine a birth mass of 320 M$_\odot$ for R136a1, which is the most massive star known to
date and the mass of which exceeds the commonly accepted upper limit of IMF significantly.
Note also that observed mass-loss rates match to within ~0.2 dex theoretical predictions \cite{Vink01} at solar metallicity and LMC metallicity for NGC 3603 and R136, respectively. 

\section{Fate of Very Massive Stars}

The fast rotating very massive stars around solar metallicity experience strong mass loss and are thus predicted to end their lives as type Ic SNe. Meanwhile for the slow rotating and non-rotating stars, the mass loss effect is less than for the rotating stars. In Table 1, we present the hydrogen surface content for 120, 150 and 200 M$_\odot$ non-rotating stars at different metallicities at the end of the main sequence. Although the models will be evolved further in order to obtain a more precise picture of the fate of these stars, if the star has lost all of its hydrogen envelope by the end of its main sequence phase, it will become a WR star and probably die as as type Ic SN. This is the case for the solar metallicity models. Stars at SMC metallicity, on the other hand, retain some of their hydrogen rich envelope at the end of the main sequence. These stars will go through the LBV phase, where the mass loss is poorly understood. They may still retain enough mass to die as PISN. For stars at LMC metallicity the fate depends strongly on the mass loss and the initial mass of the star. The fate of very massive stars at lower metallicities will be investigated in a future work.

\begin{table}
\center
\begin{tabular}{ | l | c | c | c | }
\hline
  M$_\odot$/Z & SMC (Z=0.002)  & LMC (Z=0.006)       & MW (Z=0.014) \\
  \hline
  120 & 0.75 &   0.40      & < 0.05\\
  150 & 0.62 &   0.32      & < 0.05\\
  200 & 0.42 & < 0.05     & < 0.05\\
  \hline
\end{tabular}
\label{table1}
\caption{Surface of hydrogen content predicted at end of the main sequence for the 120, 150 and 200 M$_\odot$ non-rotating stars.}
\end{table}

\acknowledgments{N.Yusof acknowledges financial support by the Ministry
of Higher Education and University of Malaya under Higher
Education Academic Training Scheme and the Commonwealth
Scholarship Commission for the Split-Site PhD 2010-2011 programme
tenable at Keele University. R. Hirschi acknowledges support from the World Premier International Research 
Center Initiative (WPI Initiative), MEXT, Japan. }

\end{document}